\documentclass[preprint,
eqsecnum,
amsmath,amssymb, aps,
]{revtex4}

\usepackage[dvipsnames]{xcolor}
\usepackage{epsfig}
\usepackage{amsfonts}
\usepackage{amsmath}
\usepackage{wasysym}
\usepackage{amssymb}
\usepackage{bm}
\usepackage{float}
\usepackage{enumitem}
\usepackage{rotating}

\usepackage{pdflscape}
\usepackage{graphicx}
\usepackage{ulem}

\usepackage[utf8]{inputenc}
\usepackage{bbm} 				
\usepackage{epstopdf}				
\usepackage{verbatim}    			
\usepackage{sidecap}                
\usepackage{indentfirst}

\newcommand{\be}{\begin{equation}}
\newcommand{\ee}{\end{equation}}
\newcommand{\bea}{\begin{eqnarray}}
\newcommand{\eea}{\end{eqnarray}}

\definecolor{armygreen}{rgb}{0.0, 0.5, 0.0}

\begin{document}

\title{Structural Boundary State Transitions in Turbulent Pipe Flow}

\author{L. Moriconi\footnote{moriconi@if.ufrj.br} and G. Saisse\footnote{giovannisaisse@pos.if.ufrj.br}}
\affiliation{Instituto de F\'\i sica, Universidade Federal do Rio de Janeiro, Av. Athos da Silveira Ramos 149, CEP: 21941-909, Rio de Janeiro, RJ, Brazil}


\begin{abstract}
Extensive optical measurements of canonical turbulent pipe flows have revealed the existence of structural boundary states (SBSs) -- near-wall low-speed streaks strongly correlated with pairs of counter-rotating quasi-streamwise vortices. In this study, we investigate the number fluctuations of these structures within the framework of statistical mechanics. Specifically, we introduce reduced degrees of freedom to model the low-speed streaks as a dilute lattice gas of hard-core particles. The Metropolis stochastic evolution provides, furthermore, a simple yet effective two-parameter model for describing SBS transitions. The lattice gas approach enables us to derive both the probability of SBS occurrence and the peculiar self-similar correlations that these structures exhibit along the streamwise direction. Our findings give additional support to the idea that the statistically stationary regimes of wall-bounded turbulent flows can be understood as Markov chains of coherent dynamical states in reduced-dimensional phase spaces.
\end{abstract}


\maketitle

\section{Introduction}

Since the landmark experiments of Osborne Reynolds about 150 years ago \cite{reynolds1}, turbulent pipe flows in their numerous variations have attracted significant interest as a fundamental subject of research. The past few decades have, in particular, been a period of remarkable advances, including (i) deeper analyses of boundary-layer scaling laws \cite{smits_etal}, (ii) the experimental finding of structural boundary states (SBSs) \cite{hof_etal, schneider_etal,dennis_sogaro, jackel1_etal, jackel2_etal}, which, upon conditioned averages, closely resemble low Reynolds-number traveling waves \cite{faisst_eck,wedin_kers,pringle_kersell}, and (iii) the elucidation of the subcritical laminar-turbulent transition within the framework of the directed percolation model of out-of-equilibrium statistical mechanics \cite{lemoult_etal}.

In short, SBSs -- our main focus in this work -- are streamwise-localized flow configurations characterized by a given number of extended wall-attached low-speed streaks, see {\hbox{Fig. 1a}} for an illustration. Such spots of relatively reduced fluid speed are typically sided by quasi-streamwise counter-rotating vortices, the legs of hairpin vortical structures. The topology of their transverse velocity field is evidenced when averages are taken over configurations that have a fixed number of low-speed streaks \cite{dennis_sogaro, jackel1_etal}.

From ensembles of sequential PIV snapshots, taken at fixed position along the pipe, one finds that the number of cross-sectional low-speed streaks, $k(t)$, keeps changing in time as a stationary stochastic process {\hbox{${\cal{S}} = \{ ..., k(t - \Delta), k(t), k(t+\Delta), ...  \}$}} \cite{jackel2_etal}. Of course, time is here discretized, as implied by the finite sampling rate $\Delta^{-1}$ of the PIV apparatus. 

We are thus challenged to recover distinctive statistical features of ${\cal{S}}$, as directly obtained from experiments, in the most synthetic way, that is, through a theoretical description that resorts to a small number of modeling parameters. We are motivated by the success of the percolation model of the laminar-turbulent transition (a one-parameter model) \cite{lemoult_etal}, to evoke statistical mechanical ideas, now in the context of the turbulent regime of the pipe flow. As discussed below, we actually establish a considerable improvement (a two-parameter model) over the stochastic model of Ref. \cite{jackel2_etal} (an eleven-parameter model), previously formulated to deal with essentially the same phenomenological issues.

\begin{figure}[h]
\hspace{0.0cm} \includegraphics[width=0.8\textwidth]{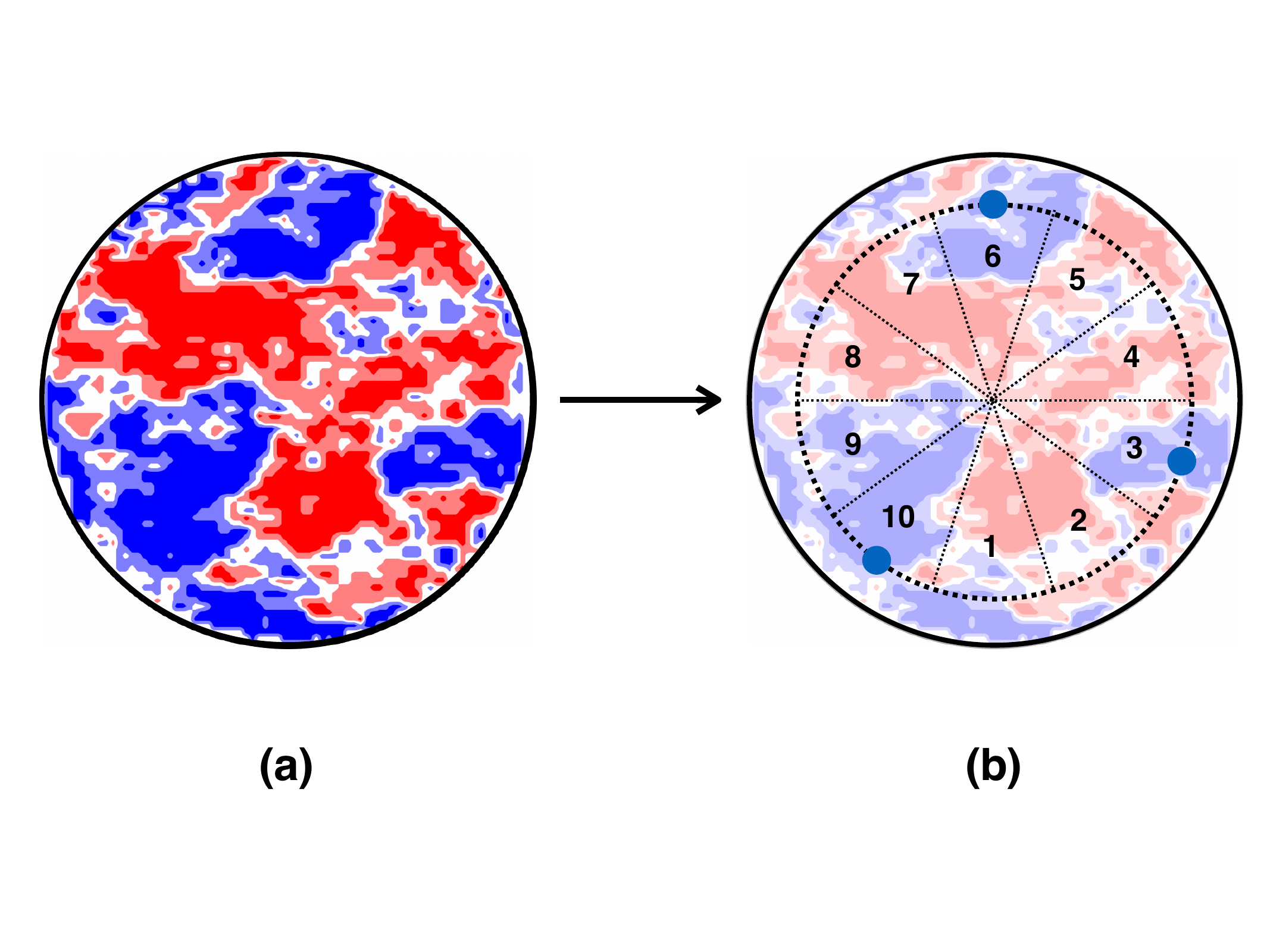}
\vspace{-1.0cm}

\caption{(a) Cross-sectional snapshot of a turbulent pipe flow ({\hbox{Re $= 24400$}}), containing three widely spread boundary low-speed streaks (blue spots), visualized with the help of particle image velocimetry (PIV) \cite{jackel1_etal}; (b) the same snapshot is mapped, in the language of streaks bits, to the microstate defined by $\sigma_3=\sigma_6=\sigma_{10} = 1$ and $\sigma_1=\sigma_2=\sigma_4=\sigma_5=\sigma_7=\sigma_8 =\sigma_9 = 0$. The dashed circle, used as a reference to the definition of SBS modes, has radius $r_0 = 0.4D$; see Eq. (\ref{ftheta}).}
\label{}
\end{figure}
\vspace{0.0cm}

This paper is organized as follows. In Sec. II, we briefly review the experimental databases and the post-processing procedures upon which our modeling analyses rely. In Sec. III, we introduce phase-space concepts for the description of the SBSs and put forward a ring lattice-gas model to account for some of their observed statistical features. 
We will mainly focus on the probability of occurrence of the SBSs and correlation functions associated to their transitions. Finally, in Sec. IV, we summarize our results and discuss modeling perspectives for additional work.

\section{structural boundary states}

In order to study the statistical properties of the stochastic process ${\cal{S}}$, we use data of water pipe flow experiments perfomed in the large pipe rig facility of the Interdisciplinary Nucleus for Fluid Dynamics (NIDF-UFRJ) \cite{jackel1_etal}. The pipe’s diameter and length are, respectively, {\hbox{$D = 15  {\hbox{~cm}}$}} and $L = 12$ m. For each of the Reynolds numbers Re = 24400 and 29000, the data consist of $2 \times 10^4$ cross-sectional snapshots of the flow, with records of the streamwise component of the turbulent velocity field taken over a uniform grid of size {\hbox{$78 \times 78$}}. The data were acquired through stereoscopic PIV, with sampling rate {\hbox{$\Delta^{-1} = 15 {\hbox{~Hz}}$}}.

SBS transitions are observed to occur for the low-speed streaks, the numbers of which mostly lie within the range {\hbox{$1 \leq k(t) \leq 10$}} \cite{jackel1_etal,jackel2_etal}. A straightforward way to get the integer numbers $k(t)$ out of the PIV snapshots is to take them as the azimuthal wavenumbers where there are power spectrum peaks of a properly defined streamwise velocity correlation function. More concretely, let $r_0$ be a reference radial distance within the log-layer (in our particular cases we take it to be $0.4D$), and $u = u(r_0,\theta,t)$ be the fluctuating part of streamwise velocity component, in polar coordinates and time instant $t$, measured in the snapshot plane. Define, then, the power spectrum function
\be
I(n,t) = \left | \int_0^{2 \pi} d \theta e^{in \theta} f(\theta,t) \right |^2 \ , \ 
\ee
where $n \in \mathbb{N}^+$ and
\be
f(\theta,t) = \int_0^{2 \pi} d \theta' u(r_0,\theta + \theta',t)
u(r_0,\theta',t) \ . \ \label{ftheta}
\ee
The number of low-speed streaks is now putatively identified to the wavenumber $k(t) = n$, which maximizes the power spectrum, that is
\be
I(n,t) > I(m,t) \ , \  {\hbox{if}}~~ m \neq n \ . \ 
\ee
For the sake of methodological consistency, we have also validated the power spectrum analysis described above through a biased image processing algorithm which localizes low-speed streaks in physical space.


\section{Lattice Gas Approach}

To begin with modeling considerations, it is useful to introduce a few concepts of statistical mechanical flavor. 
\vspace{0.2cm}

\noindent $\star$ By a SBS of {\it{mode}} $k$ we mean that there are $k$ low-speed streaks in a given PIV snapshot.
\\

\noindent $\star$ Taking into account that there are at most 10 SBS modes, the PIV cross-sectional plane is sliced into ten {\it{strike channels}}, a construction illustrated in Fig. 1b.
\\

\noindent $\star$ A given streak channel of label $1 \leq i \leq 10$ can be occupied or not by a low-speed streak. We model its occupancy by the {\it{streak bits}} $\sigma_i = 1$ or {\hbox{$\sigma_i = 0$}}.
\\

\noindent $\star$ The phase-space of SBSs has dimension $2^{10}-1$. Its {\it{microstates}} are streak bit vectors of the form
$
\mathcal{M} = \{\sigma_1,\sigma_2, \ ... \, \sigma_{10} \}
$, with $\sum_i \sigma_i > 0$.
\\

An example of a microstate associated to the SBS mode $k=3$ is shown in Fig. 1b. One could claim that both streak channels 9 and 10 are occupied. However, they belong to the same low-speed structure. The particular choice of streak bits $\sigma_9 =0$ and $\sigma_{10} =1$ follows the ambiguity-resolving prescription (implemented in the alternative image processing algorithm) of selecting the streak channel where the streamwise flow rate is most significantly reduced.

A key aspect of our analysis, which is important to highlight, is the relationship between microstates and the SBS modes. Specifically, it is clear that a given SBS mode $k$ can be microscopically realized by a number of $\binom{10}{k}$ microstates.

This distinction between microstates and SBS modes is analogous to the concept of macrostates and microstates in statistical mechanics. In statistical mechanics, a macrostate represents a broad configuration defined by macroscopic variables (like temperature or pressure), while the microstates are the specific, detailed dynamical arrangements of particles that correspond to that macrostate. 
The combinatorial factor $\binom{10}{k}$ reflects the number of microstates that can realize a particular SBS mode, just as the number of microstates corresponds to the ways a given macrostate can be realized in statistical mechanics.


We are now ready to present a ring lattice model to describe the SBS fluctuations. The streak channels are associated to the sites of a lattice gas along a ring of radius $r_0$, as depicted in Fig. 1b. Low-speed streaks are the ``particles" of this gas, which have bounded occupancy given by the streak bits $\sigma_i$.  Assuming statistical mechanical equilibrium, the general grand canonical partition function of the system can be written as
\be
Z = \sum_{\{\sigma_i\}} \exp \left [-H(\{\sigma_i\}) + \mu \sum_i \sigma_i \right ] \ , \ \label{Z}
\ee
where $\mu$ is the chemical potential of the gas and the Hamiltonian $H(\{\sigma_i\})$ is taken to depend only on the microstates. In (\ref{Z}), the summation is carried out over microstates that have at least one non-vanishing streak bit, once the SBS mode $k=0$ is never observed.

It falls outside the scope of this work to provide a first-principles derivation of the Hamiltonian $H(\{\sigma_i\})$. However, having in mind the interactions between pairs of vortex structures, which are correlated with low-speed streaks -- specifically the counter-rotating legs of hairpin vortices, as schematized in Fig. 2 -- we propose a simplified, non-local Hamiltonian that accounts for their mutual interactions, as
\begin{equation}
H(\{ \sigma_i \} ) = J \sum_{i<j} \sigma_i \sigma_j \ . \ \label{H}
\end{equation}
It is not clear, \textit{a priori}, whether $J$ is a positive or negative coupling constant. Its value depends on the specific details of the quasi-streamwise/hairpin vortex configurations and, like the chemical potential, will ultimately be determined through comparisons between analytical predictions and experimental results.

Notice that $J$ and $\mu$ are dimensionless, taken as energy parameters expressed in units of some energy scale of the flow. In principle, all we can say, at this level of modeling, is that they are Reynolds number dependent quantities. 


\begin{figure}[ht]
\hspace{0.0cm} \includegraphics[width=0.47\textwidth]{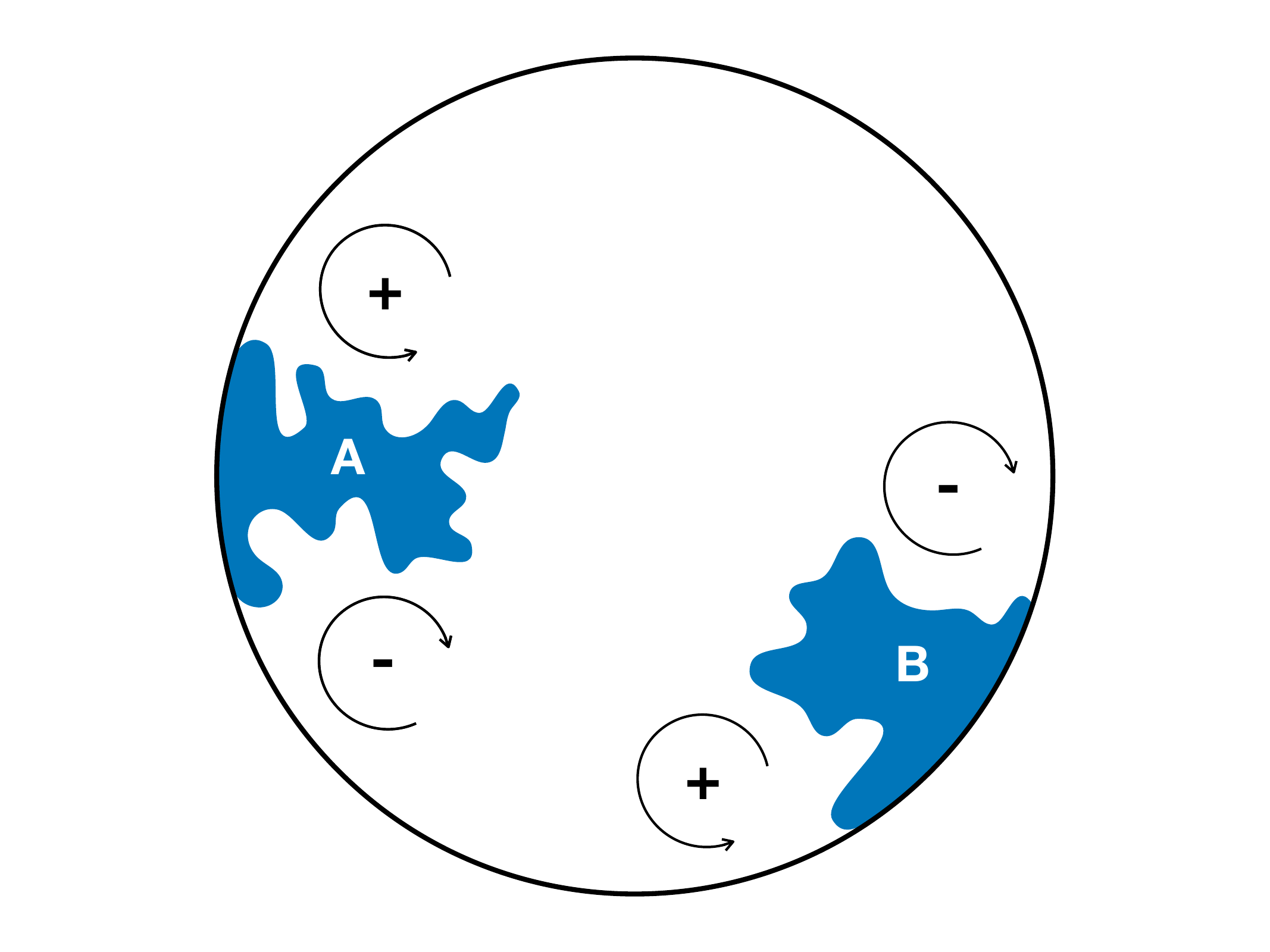}
\vspace{0.0cm}

\caption{The low-speed streaks A and B of a hypothetical SBS mode $k=2$ are flanked by pairs of counter-rotating vortices.}
\label{}
\end{figure}
\vspace{0.0cm}

\begin{figure}[t]
\hspace{0.0cm} \includegraphics[width=0.55\textwidth]{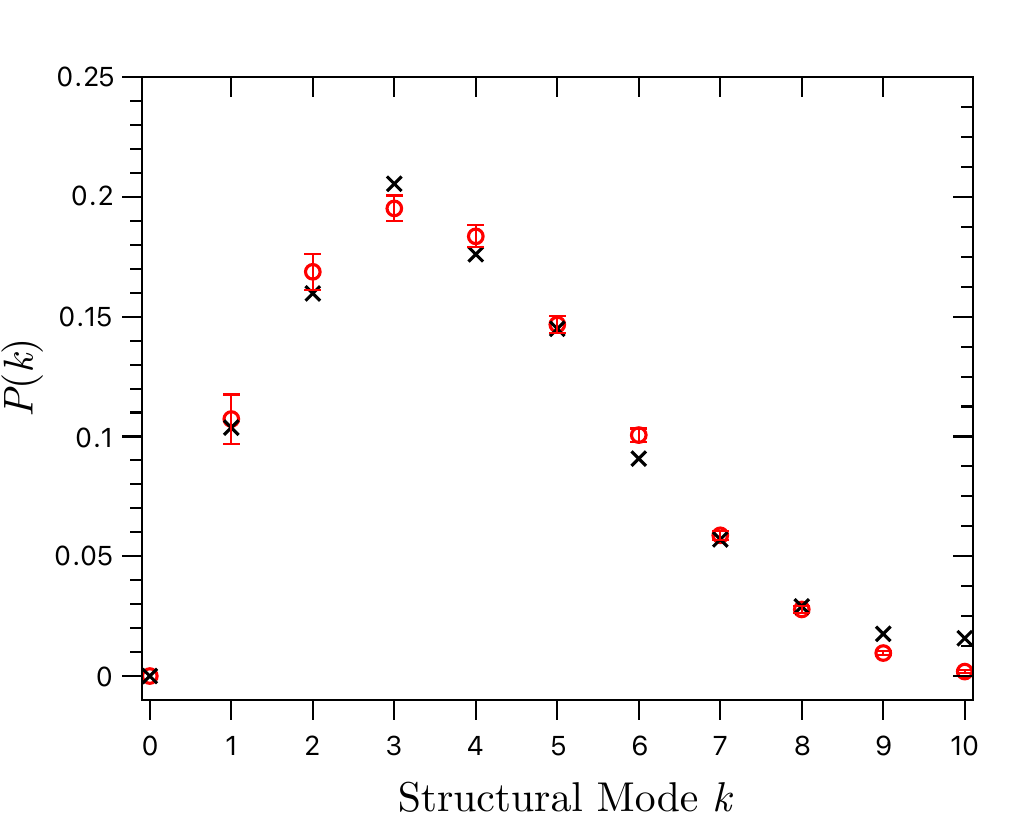}
          \put (-100,150){\makebox[0.05\linewidth][r] {\scalebox{1.2} {(a)}}}
          \put (-150,50){\makebox[0.05\linewidth][r] {\scalebox{1.2} {Re = 24400}}}

\hspace{0.0cm} \includegraphics[width=0.55\textwidth]{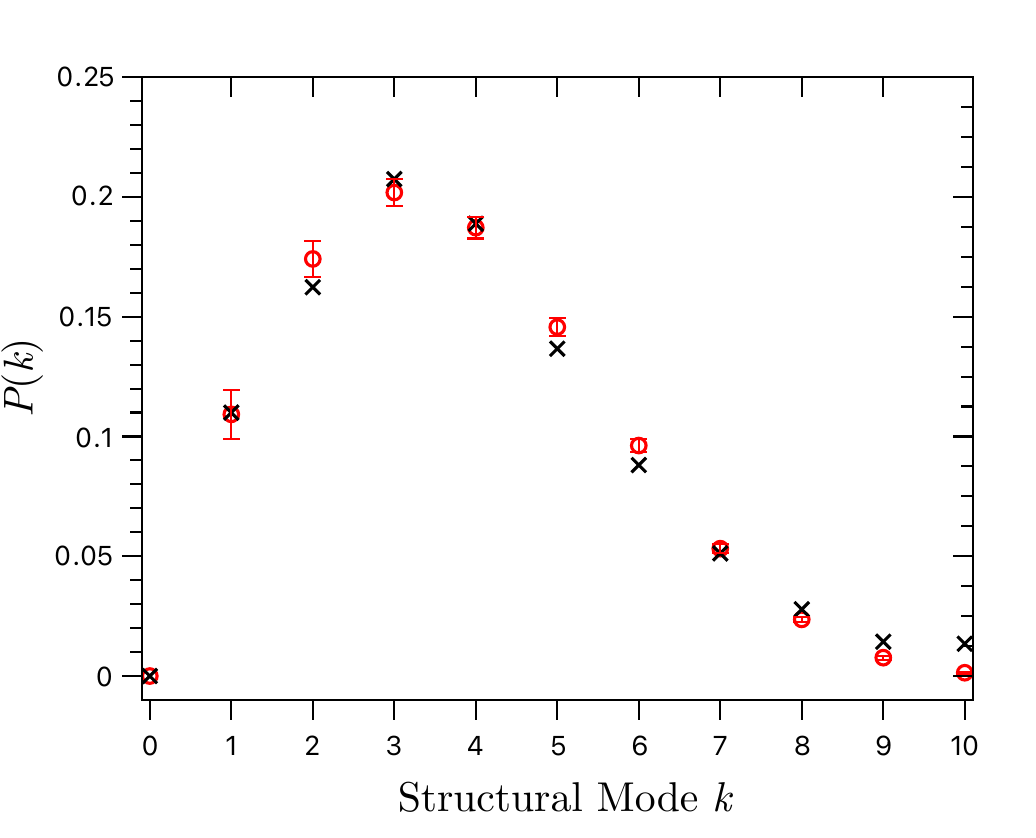}
          \put (-100,150){\makebox[0.05\linewidth][r] {\scalebox{1.2} {(b)}}}
          \put (-150,50){\makebox[0.05\linewidth][r] {\scalebox{1.2} {Re = 29000}}}
\vspace{-0.0cm}
\caption{Comparison between the empirical probability distribution of SBS modes (crosses) and the 
ones predicted by Eq. (\ref{P(k)}) (red circles), with modeling parameters (a) $J = -2.14 \times 10^{-1}$, 
$\mu = -1.26$ and (b) $J = -2.02 \times 10^{-1}$, 
$\mu = -1.24$. The error bars for the predicted probabilities are evaluated from Monte Carlo analysis, as described in the text.}
\label{}
\end{figure}

From (\ref{Z}) and (\ref{H}) we get, without any difficulty, the probability for the occurrence of the SBS mode $k$, as
\be
P(k) = \frac{1}{Z} \binom{10}{k} \exp \left [ - \frac{J}{2} k^2 + \left ( \mu + \frac{J}{2} \right )
k \right ] \ , \ \label{P(k)}
\ee
where  $1 \leq k \leq 10$ (recall that $P(0) = 0$). Optimal values of $J$ and $\mu$ can be determined from the empirical probabilities, directly obtained from the experimental databases, through the evaluation of L2 losses. Very reasonable results are reported in Fig. 3. There, we show, besides the analytical results, the error bars predicted by fluctuations of 500 Monte Carlo samples, each containing $2 \times 10^4$ SBS modes, the same size of the PIV snapshot sets.

We have implemented the Monte Carlo approach using both the Metropolis and Glauber algorithms for performance comparisons \cite{binder_heermann}. Both methods yield essentially equivalent ensemble averages, with the Metropolis algorithm offering a slight advantage in accuracy, which is why we choose to adopt it.

A microscopic state $\{ \sigma_i(n) \}$ at time $n$ (integer) is proposed to evolve via the Metropolis algorithm to a new microstate $\{ \sigma_i(n+1) \}$ at time $n+1$. This new microstate is randomly selected from the set of $2^{10} - 1$ possible configurations. Defining now the grand canonical energy, which includes chemical potential contributions,
\be
H_\mu(\{\sigma_i\}) = J \sum_{i<j} 
\sigma_i \sigma_j -\mu \sum_i \sigma_i \ , \ 
\ee
its virtual change along a Monte Carlo step is 
\be
\Delta (H_\mu) = J \sum_{i<j} 
\sigma_i(n+1) \sigma_j(n+1) -  \sigma_i(n) \sigma_j(n) ] - \mu \sum_i [\sigma_i(n+1) -\sigma_j(n)] \ . \ 
\ee
We accept the proposed change of microstates, as usual, if $\Delta H_\mu \leq 0$. Otherwise, the modification is accepted with probability $\exp[-\Delta ( H_\mu )]$.

\begin{figure}[t]
\hspace{0.0cm} \includegraphics[width=0.55\textwidth]{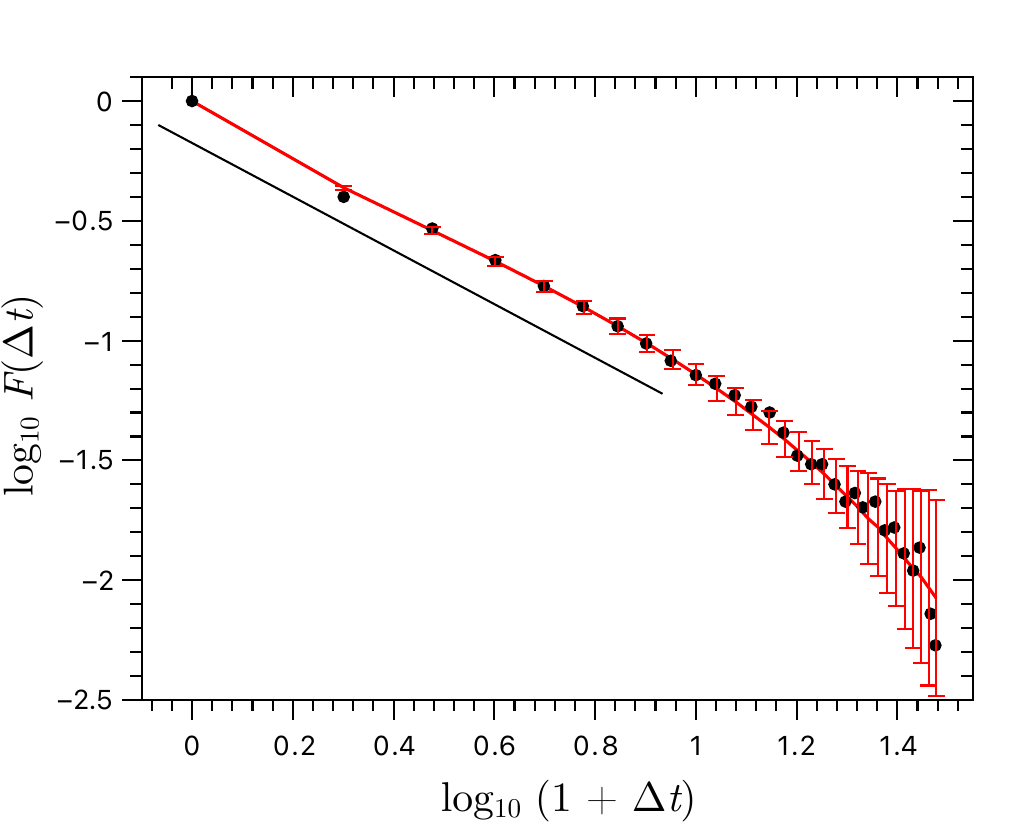}
          \put (-100,150){\makebox[0.05\linewidth][r]
          {\scalebox{1.2} {(a)}}}
          \put (-150,50){\makebox[0.05\linewidth][r] {\scalebox{1.2} {Re = 24400}}}

\hspace{0.0cm} \includegraphics[width=0.55\textwidth]{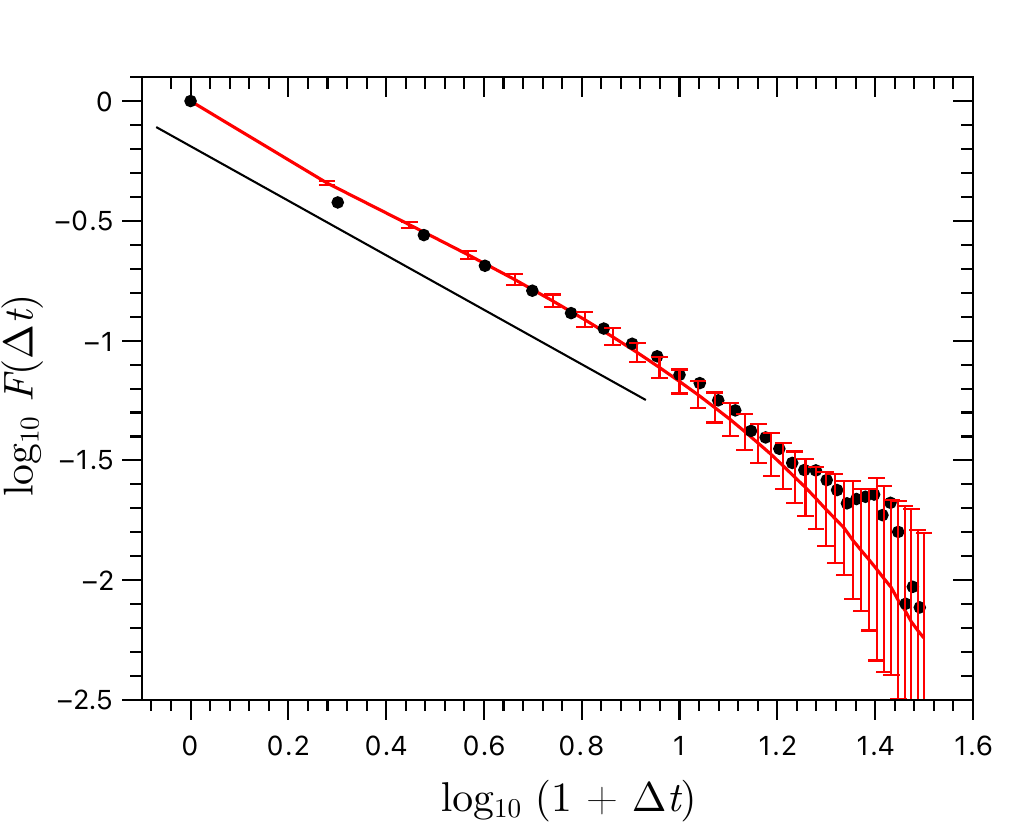}
          \put (-100,150){\makebox[0.05\linewidth][r] {\scalebox{1.2} {(b)}}}
          \put (-150,50){\makebox[0.05\linewidth][r] {\scalebox{1.2} {Re = 29000}}}
\vspace{-0.0cm}
\caption{Measurements and modeling of the mode correlation function (\ref{F}). Dots refer to experimental data, while the red solid lines and error bars refer to the Monte Carlo predictions taken over $500$ samples, each containing $2 \times 10^4$ microstates. Straight lines: (a) $F(\Delta t) = (1+\Delta t)^{-1.12}$, (b) $F(\Delta t) = (1+\Delta t)^{-1.14}$, both are displaced downward to improve visualization. The time rescaling factors are, respectively, $\gamma = 1.0$ 
and {\hbox{$\gamma = 0.9$.}}}
\label{}
\end{figure}

A more demanding test of the lattice gas model of SBS fluctuations, which assumes that the Metropolis evolution algorithm captures dynamical features of the sampled stochastic process, is related to the evaluation of the mode correlation function
\be
F(t'-t) = \frac{\langle \delta[ \sum_i \sigma_i(t), \sum_i \sigma_i(t')  ] \rangle - 
\sum_k 
\langle \delta[ \sum_i \sigma_i(t), k ] \rangle^2}{1 - \sum_k 
\langle \delta[ \sum_i \sigma_i(t), k ] \rangle^2} \ , \ \label{F}
\ee
previously introduced in \cite{jackel2_etal}. Above, $\delta [\cdot ,\cdot]$ stands for the Kronecker delta function. Its clear that $F(t'-t)$ is normalized to $F(0)=1$ and that 
$F(t'-t)$ vanishes for large times. 

The importance of (\ref{F}) is connected with the idea of SBS mode recurrence. We define two microstates at times $t$ and $t'>t$,
to be {\it{mode recurrent}} if
\be
\sum_i \sigma_i(t) = \sum_i \sigma_i(t')   \ . \ 
\ee
 It is not difficult to show that $F(t'-t)=1$ if in all realizations of the stationary stochastic process the pairs of microstates at times $t$ and $t'$ are mode recurrent. In contrast, if microstates are completely uncorrelated, it follows that $F(t'-t)=0$.

The motivation for studying (\ref{F}) stems, thus, from the relevant role of quasi-periodic orbits in the theory of chaotic dynamical systems and turbulent flows, where one may work with dimensionally reduced phase spaces associated with coherent structures \cite{gibson_etal,moehlis_etal,budanur_etal}. In the language of stochastic processes, recurrence can be characterized by a slowly decaying profile of $F(t'-t)$ within the integral time scale of the flow \cite{jackel2_etal}.

\begin{figure}[t]
\hspace{0.0cm} \includegraphics[width=0.55\textwidth]{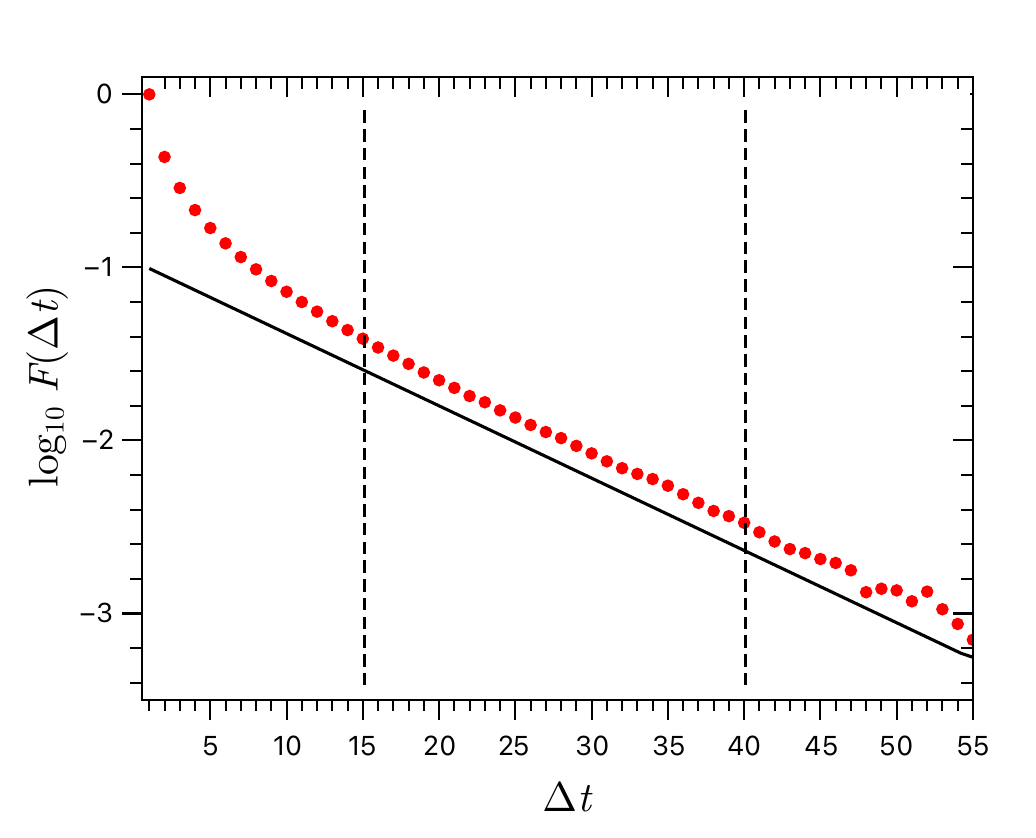}
          \put (-100,150){\makebox[0.05\linewidth][r]
          {\scalebox{1.2} {(a)}}}
          \put (-100,50){\makebox[0.05\linewidth][r] {\scalebox{1.2} {Re = 24400}}}

\hspace{0.0cm} \includegraphics[width=0.55\textwidth]{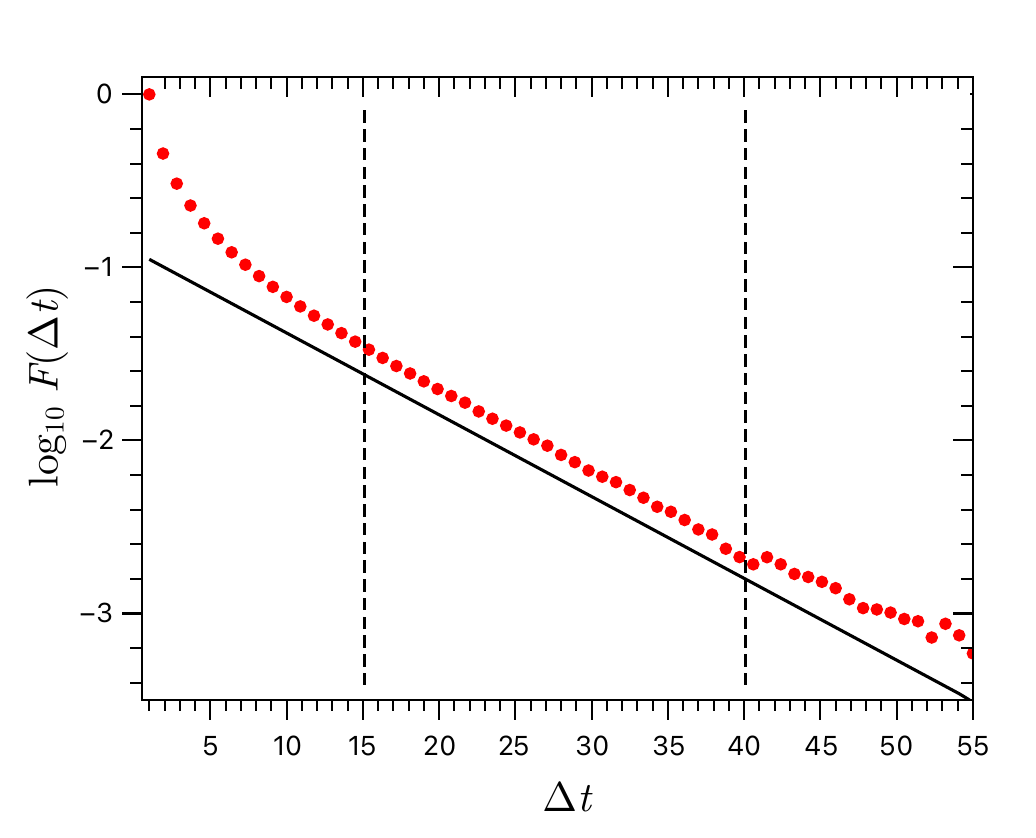}
          \put (-100,150){\makebox[0.05\linewidth][r] {\scalebox{1.2} {(b)}}}
          \put (-100,50){\makebox[0.05\linewidth][r] {\scalebox{1.2} {Re = 29000}}}
\vspace{-0.0cm}
\caption{Dots represent the Monte Carlo simulations of the mode correlation function (\ref{F}) in a mono-log plot. The solid straight lines are linear fits taken in the interval $15 \leq \Delta t \leq 40$: (a) $\log_{10} F(\Delta t) = 
-0.81 - 0.042 \Delta t$, (b) $\log_{10} F(\Delta t) = 
-0.76 - 0.047 \Delta t$. Both are displaced downward to improve visualization.}
\label{}
\end{figure}

Our task, now, is to model $F(t'-t)$ with the stochastic time series produced by the Metropolis simulation protocol. While the natural stochastic process is recorded at discrete time intervals $\Delta$, the time series generated from the Monte Carlo simulation employs a unit time step. Since these time scales differ between the two processes, it is necessary to adjust the time scales to ensure an accurate representation of the natural process through the Monte Carlo simulation. Therefore, for the sake of systematic comparisons with the Monte Carlo simulations, the time measurements of the natural processes are expressed in ``snapshot time units" $\Delta$. To account for further phenomenological differences of time parametrization, we introduce a time rescaling factor $\gamma$ that modifies the time interval of the Monte Carlo-generated time series. This factor effectively rescales the simulation’s time steps, aligning them with the time intervals $\Delta$ of the natural process. 

When analyzing the correlation functions of both processes, this time scale adjustment will result, in principle, in a horizontal shift of the curves in a log-log plot. This shift enables a meaningful comparison between the correlation structures of the natural and simulated processes, despite their original definitions in arbitrary time units.

As we can see from Fig. 4, the model is able to reproduce, very sharply, the correlation function (\ref{F}). For both of the investigated Reynolds numbers, we find that up to the physical time lag $\Delta t \approx 15$,
\be
F(\Delta t) \approx \frac{1}{(1+ \Delta t)^{1.1}} \ . \
\ee
A noticeable aspect of the Monte-Carlo results is that the error bars suddenly increase for $\Delta > 15$, which seems to be the case for the real data as well.

The correlation functions exhibit a crossover around $\Delta t \approx 15$ (= 1 second) from their slower, power law decay, to a much faster decaying exponential behavior, as indicated in Fig. 5. It is worth emphasizing that the crossover time is of the order of the integral time scale of the flow $D^2/\nu {\hbox{Re}}$, where $\nu \approx 10^{-6}$ is the kinematic viscosity of water.

\section{Conclusions}

We have developed a statistical mechanical model for turbulent pipe flow, considering not only its stationary statistical regime but also 
mode fluctuations (via Monte Carlo simulations) associated to SBS transitions.

As a central result of this work, it is shown that a Markovian description of transitions between structural boundary microstates can effectively characterize some dynamical features of turbulent pipe flow. This approach opens up intriguing possibilities for exploring complex phenomena in non-Newtonian or multiphase fluids, magnetohydrodynamics, and other related settings.

A first-principles derivation of the Hamiltonian (\ref{H}) would provide deeper insights into the underlying mechanisms governing the flow. It is likely that refinements of the lattice gas model will incorporate random couplings and alternative non-local interactions in the Hamiltonian. Expanding the experimental dataset and conducting more comprehensive statistical analyses will be crucial in strengthening our understanding of the SBS stochastic dynamics.

In summary, while the current model offers a robust representation of some aspects of turbulent pipe flow, there is still great room for improvement and extension of the theory, with the expectation of valuable insights into the broader applications of statistical mechanics in fluid turbulence.
\vspace{0.3cm}

\leftline{{\it{Acknowledgments}}}
\vspace{0.3cm}

The authors thank the Conselho Nacional de Desenvolvimento Científico e Tecnológico (CNPq) and the Coordenação de Aperfeiçoamento de Pessoal de Nível Superior (CAPES) for partial support.

\end{document}